\documentclass{pasj00}
\draft
\onecolumn

\usepackage[dvips]{color}

\begin{document}
\SetRunningHead{Y. Suwa et al.}{Magnetorotational Collapse of Pop III}
\Received{2006/10/12}
\Accepted{2007/4/11}

\title{Magnetorotational Collapse of Population III Stars
 }

\author{
Yudai \textsc{Suwa}\altaffilmark{1},
Tomoya \textsc{Takiwaki}\altaffilmark{1},
Kei \textsc{Kotake}\altaffilmark{2}, and, 
Katsuhiko \textsc{Sato}\altaffilmark{1,3}
}
\altaffiltext{1}
{Department of Physics, School of Science,
The University of Tokyo, Tokyo 113-0033}
\altaffiltext{2}
{National Astronomical Observatory of Japan, Mitaka, Tokyo 181-8588,Japan}
\altaffiltext{3}
{Research Center for the Early Universe,
School of Science, the University of Tokyo,7-3-1 Hongo,\\
Bunkyo-ku, Tokyo 113-0033, Japan}

\email{suwa@utap.phys.s.u-tokyo.ac.jp}


\maketitle

\begin{abstract}
We perform a series of two-dimensional magnetorotational core-collapse simulations of 
Pop III stars. Changing the initial distributions of rotation and magnetic 
fields prior to collapse in a parametric manner, 
we compute 19 models. By so doing, 
we systematically 
investigate how rotation and magnetic fields affect 
the collapse dynamics and explore how the properties of 
the black-hole formations and neutrino emissions could be affected.
As for the microphysics, we employ a realistic equation of state and 
approximate the neutrino transfer by a multiflavour leakage scheme. With these 
computations,
we find that the jet-like explosions are obtained by the 
magneto-driven shock waves if the initial magnetic field is as large as 
$10^{12}$G. 
We point out that the black-hole masses at the formation decrease with 
 the initial field strength, on the other hand, increase with the initial 
rotation rates.
As for the neutrino properties, we point out that 
the degree of the differential rotation plays an important role to 
determine which species of the neutrino luminosity is more dominant 
than the others. Furthermore, we find that the stronger magnetic fields make 
the peak neutrino luminosities smaller, because the magnetic 
pressure acts to halt the collapse in the central regions, leading 
to the suppression of the releasable gravitational binding energies.
\end{abstract}

\section{INTRODUCTION}
Great attention has been paid to Population III, the first stars to form 
in the universe, because they are related to many unsettled problems in 
cosmology and the stellar physics. 
The Population III 
(Pop III) stars ionize and enrich the metallicity of the intergalactic medium 
and thus provide important clues to the subsequent star formation history 
(for reviews, see e.g., \cite{bark01,brom04,glov05}).
Pop III stars are also important for the understanding of the chemical 
evolution history. 
 Recent discovery of hyper metal poor stars such as 
HE 0107-5240 \citep{chri02} and HE 1327-2326 \citep{freb05} has given us good opportunities to investigate 
the nucleosynthesis in Pop III stars \citep{heger02,umed02,umed03,daig04,iwam05}.
Gamma-ray bursts at very high 
redshift are pointed out to be accompanied 
by the gravitational collapse of Pop III stars \citep{schn02,brom06}.    
The {\it Swift} satellite, which is now running \footnote{See http://swift.gsfc.nasa.gov},
is expected to directly detect the
 Pop III stars accompanied by 
the high-z gamma-ray bursts. 

The evolutions of Pop III stars have 
been also  studied for long. 
From their studies, Pop III stars are predicted to have been 
predominantly very massive with $M\gtrsim 100M_\odot$ 
(\cite{naka01, abel02, brom02b}, see references therein).
Massive stars in the range of $100 M_{\odot} \lesssim M 
\lesssim 260M_\odot$ 
encounter the electron-positron pair instability 
during their evolution. This instability sets off explosive oxygen burning, 
and if the burning provides enough energy to reverse the collapse, 
the stars are thought to become pair instability supernovae \citep{bond84, fryer01,heger02}, 
whose detectability has been recently reported 
\citep{scan05,wein05}.
More massive stars, which also encounter pair-instability, are so tightly bound and the fusion of oxygen 
is unable to reverse infall. 
Such stars are thought to collapse to black holes (BHs) finally 
\citep{bond84,fryer01}, which we 
pay attention to in this paper. 

So far there have been a few hydrodynamic simulations studying the 
gravitational-collapse of the BH forming Pop III stars.
In the two-dimensional, gray neutrino transport simulations 
by \citet{fryer01}, 
they investigated the collapse of a rotating Pop III star of $300 M_{\odot}$, 
 leading to the BH formation. They discussed the effects of 
rotation on the emitted neutrino luminosities, gravitational waves, and 
furthermore, the possibility of such stars to be the gamma-ray bursts.
In their Newtonian study, the central BH was excised and treated 
as an absorbing boundary after the formation. Although such simplification is 
not easy to be validated, they followed the dynamics long after the 
formation of the BH and obtained many findings.
More recently, \citet{naka06} performed 
one-dimensional, but, general relativistic simulations 
in the range of $100 \sim 10000 M_{\odot}$, 
in which the state-of-the-art neutrino physics are taken into account. 
Their detailed calculations revealed the properties of the emergent neutrino 
spectrum, and based on that, they discussed the detectability of such 
neutrinos as the supernova relic neutrino background (see also \cite{ando04,iocco05}).
They successfully saw the formation of the apparent horizon, however, the dynamics in the later phases was not referred.

 In this paper we study the magnetorotational collapse 
of Pop III stars by performing the two-dimensional magnetohydrodynamic (MHD) simulations
(see, also, \cite{aki03,kota04a,kota04b,taki04,yama04,arde05,sawa05,ober06}, 
for MHD computations of core-collapse supernovae, and \cite{kotarev} for a review).
As for the microphysics, we employ a realistic equation 
of state based on the relativistic mean field theory and 
 take into account the neutrino cooling by a multiflavor leakage scheme, 
in which state-of-the-art reactions of neutrinos are included. In our Newtonian simulations,
 the formation of the BHs is ascribed to a certain condition, 
and after the formation, the central region is excised and treated as an 
absorbing boundary in order to follow the dynamics later on.
Since the distributions of rotation and magnetic fields in the progenitors 
of Pop III stars are highly uncertain, we change them in a parametric 
manner and systematically investigate how rotation and magnetic fields 
affect the dynamics.
We also explore how the natures of explosions, the properties of the BHs 
and neutrino luminosities could be affected due to the incursion of 
the rotation and magnetic fields.

This paper is organized as follows.
In \S2, we describe the numerical methods and the initial conditions.
In \S3, we present the results. We give a summary and discussion in \S4.

\section{METHOD}
\subsection{Basic Equations}
The basic evolution equations are written as follows,
\begin{equation}
\frac{\mathrm{d}\rho}{\mathrm{d}t}+\rho\nabla\cdot\mathbf{v}=0,
\end{equation}
\begin{equation}
\rho \frac{\mathrm{d}\mathbf{v}}{\mathrm{d}t}=-\nabla P -\rho \nabla \Phi
+\frac{1}{4\pi}\left(\mathbf{\nabla}\times\mathbf{B}\right)\times\mathbf{B},
\end{equation}
\begin{equation}
\rho\frac{\mathrm{d}}{\mathrm{d}t}\left(\frac{e}{\rho}\right)=-P\nabla \cdot \mathbf{v}-L_{\nu}\label{cool},
\end{equation}
\begin{equation}
\frac{\partial\mathbf{B}}{\partial t}= \mathbf{\nabla} \times \left(\mathbf{v}\times\mathbf{B}\right)\label{eq:1},
\end{equation}
\begin{equation}
\bigtriangleup{\Phi} = 4\pi G \rho,
\end{equation}
where $\rho,P,\mathbf{v},e,\Phi,\mathbf{B},L_{\nu}$, 
$\frac{\mathrm{d}}{\mathrm{d}t}$, are the mass density, the gas 
pressure including the radiation pressure from neutrino's, the fluid 
velocity, the internal energy density, the gravitational potential, 
the magnetic field, the neutrino cooling rate, and Lagrange 
derivative, respectively.
In our 2D calculations, axial symmetry and reflection symmetry across 
the equatorial plane are assumed.
Spherical coordinates $(r,\theta)$ are employed with logarithmic 
zoning in the radial direction and regular zoning in $\theta$.
One quadrant of the meridian section is covered with 300 
($r$)$\times$ 30 ($\theta$) mesh points. 
The minimum and maximum mesh spacings are 2 km and 60 km, respectively.
We also calculated some models with 60 angular mesh points, however, 
any significant difference was obtained.
Therefore, we will report in the following the results obtained from the models 
with 30 angular mesh points.
We employed the ZEUS-2D code \citep{ston92} as a base and added major 
changes to include the microphysics.
First we added an equation for electron fraction to treat electron 
captures and neutrino transport by the so-called leakage scheme \citep{kota03}.
Furthermore, we extend the scheme to include all 6 species of neutrino 
($\nu_e,\bar\nu_e,\nu_X$), which is indispensable for the computations of 
the Pop III stars.
Here $\nu_X$ means $\nu_\mu,\bar\nu_\mu,\nu_\tau$ and $\bar\nu_\tau$.
As for the reactions of $\nu_X$, 
pair, photo, and plasma processes
are included using the rates by \citet{itoh89}.
The $L_{\nu}$, in Eq. (\ref{cool}) is the cooling rate of the relevant neutrino reactions
(see \cite{taki06}, for details).
As for the equation of state, we have incorporated the tabulated one based on 
relativistic mean field theory instead of the ideal gas EOS 
assumed in the original code \citep{shen98}.

\subsection{Initial Models and Boundary Condition}
In this paper, we set the mass of the Pop III star to be $300 M_{\odot}$.
This is consistent with the recent simulations of the star-formation phenomena 
in a metal free environment, providing an initial mass function peaked 
at masses $100 - 300 M_{\odot}$ (see, e.g., \cite{naka01}).
We choose the value because we do not treat 
the nuclear-powered pair instability supernovae ($M \lesssim 260 M_{\odot}$) and, for convenience, for the comparison with the previous study, which employed the same 
stellar mass \citep{fryer01}.  
 
 We start the collapse simulations of $180M_\odot$ core of the $300M_\odot$ 
star. The core, which is the initial condition of our simulations, is produced 
in the following way. According to the prescription 
in \citet{bond84}, we set the polytropic index of the core to $n=3$ and 
assume that the core is isentropic of $\sim10k_B$ per nucleon \citep{fryer01}
 with the constant electron fraction of $Y_e=0.5$.
We adjust central density to $5\times10^6~{\rm g}~{\rm cm}^{-3}$, by which the 
temperature of the central regions become high enough to photodisintegrate the 
iron ($\sim 5\times10^9$K), thus initiating the collapse.
Given central density, the distribution of electron fraction, and entropy, 
we construct numerically the hydrostatic structures of 
the core.

Since we know little of the angular momentum distributions in the cores 
of Pop III stars (see, however, \cite{fryer01}), we add the following 
rotation profiles in a parametric manner to the non-rotating core mentioned 
above.
We assume the cylindrical rotation of the core and change 
the degree of differential rotation in the following two ways. 
\begin{enumerate}
	\item As for the differential rotation models, we assume the 
	following distribution of the initial angular velocity,
	  \begin{equation}
	\Omega(X,Z)=\Omega_{0}\frac{X_{0}^2}{X^2+X_{0}^2}\frac{Z_{0}^4}{Z^4+Z_{0}^4}, \label{dif. rot.}
\end{equation}
	where $\Omega$ is the angular velocity and $\Omega_0$ is the model constant. $X$ and $Z$ denote distance 
from rotational axis and the equatorial plane, respectively. 
We adopt the value of parameters, $X_0$ and $Z_0$, as  $2\times10^8 
\mathrm{cm},2\times10^9\mathrm{cm}$, respectively.
Since the radius of the outer edge of the core is taken to be as large as   
$3.5\times 10^9 \mathrm{cm}$, the above profile represents that the cores 
rotate strongly differentially.
	\item As for the rigid rotation models, the initial angular velocity is 
given by, 
	\begin{equation}
	\Omega(X,Z)=\Omega_0.
	\end{equation}
\end{enumerate}

\begin{figure}[htbp]
\begin{center}
\FigureFile(80mm,80mm){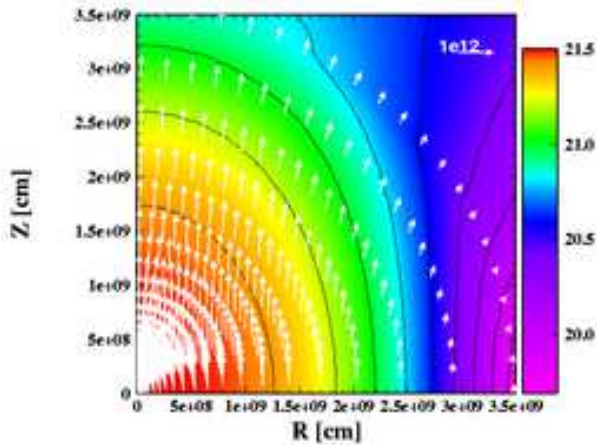}
\end{center}
  \caption{The configuration of the initial magnetic fields. 
Note that $B_0=10^{12}$G for this figure. 
The arrows represent the vector of the poloidal magnetic fields. 
The contour shows the logarithm of the magnetic pressure ($:B^2/8\pi)$.}
 \label{fig:ini_mag}
\end{figure}

As for the initial configuration of the magnetic fields, we assume that 
 the field is nearly uniform and parallel to the rotational axis in the core 
and dipolar outside (see Figure \ref{fig:ini_mag}). 
For the purpose, we consider the following effective vector potential,
\begin{equation}
A_r=A_\theta=0,
\end{equation}
\begin{equation}
 A_\phi=\frac{B_0}{2}\frac{r_0^3}{r^3+r_0^3}r\sin\theta,
\end{equation}
where $A_{r,\theta,\phi}$ is the vector potential in the $r,\theta,\phi$ 
direction, respectively,  $r$ is the radius, $r_0$ is the radius of the core, and 
$B_0$ is the model constant.
In this study, we adopt the value of $r_0$ as $3.5\times10^9$ cm.
This vector potential can produce the uniform magnetic fields when $r$ 
is small compared with $r_0$, and the dipole magnetic fields for vice versa.
We set the outflow boundary conditions for the magnetic fields at the 
outer boundary of the calculated regions.
It is noted that this is a far better way than the loop current method 
for constructing the dipole magnetic fields \citep{lebl70}, 
because our method produces no divergence of the magnetic fields 
near the loop current.

Changing the initial rotational and magnetic energies by 
varying the values of $\Omega_0$ and $B_0$, 
we compute 19 models in this paper, namely, one spherical and 18 
magnetorotational models.
In Table \ref{tab:model}, we summarize the differences of the initial models. 
Note that the models are named after this combination, with the first letters, B12, B11, B10, indicating the strength of initial magnetic field, the following letter, TW1, TW2, TW4 indicating the 
initial $T/|W|$ and final capital letter D or R representing the initial rotational law (D: Differential rotation, R: Rigid rotation). Note that $T/|W|$ represents the 
ratio of the rotational to the gravitational energy.

\begin{table*}[htbp]
\begin{center}
\caption{Models and Parameters.\footnotemark[$*$]
}
\label{tab:model}
\begin{tabular}{|c|ccc|}
\hline\hline
 &\multicolumn{3}{c|}{$T/|W|$} \\
\cline{2-4}
$B_{0}$ & 1\% &2\%  &4\%\\
\hline
$10^{10}$G    & B10TW1\{D,R\} & B10TW2\{D,R\} & B10TW4\{D,R\}  \\
$10^{11}$G    & B11TW1\{D,R\} & B11TW2\{D,R\} & B11TW4\{D,R\}  \\
$10^{12}$G    & B12TW1\{D,R\} & B12TW2\{D,R\} & B12TW4\{D,R\}  \\
\hline
\multicolumn{1}{@{}l@{}}{\hbox to 0pt{\parbox{95mm}{\footnotesize
    \footnotemark[$*$] 
    This table shows the name of the models. In the table they are 
    labeled by the strength of the initial magnetic field and rotation. 
    $T/|W|$ represents the ratio of the rotational to the gravitational 
    energy. $B_{0}$ represents the strength of the initial magnetic field.
}\hss}}
\end{tabular}
\end{center}
\end{table*}

In this paper we assume a BH is formed when the condition 
$\frac{6Gm(r)}{c^2}>r$ is satisfied, where $c,G,m(r)$ are the speed of light, the gravitational constant and the mass 
coordinate, respectively.
This condition means that we assume that fluids cannot escape from the inner region below the radius of the marginally stable orbit of a Schwarzschild BH. 
When this condition is satisfied, we excise the region inside the radius 
calculated and then treat it as an absorbing boundary. Afterwards, we enlarge 
the boundary of the excised region to take into account the growth of the mass 
infalling into the central region.  
Although it is not accurate at all to refer the central region as the BH, we cling to the simplification in this paper in order to follow and see 
the dynamics later on.

\section{RESULT}\label{sec:result}

\subsection{Spherical Collapse}

First of all, we briefly describe the hydrodynamic features of 
spherical collapse as a baseline for the MHD models mentioned later.
Note in the following that by ``massive stars'', we mean the stars of
$\approx O(10) M_{\odot}$ with the initial composition of the solar 
metallicity, which are considered to explode as supernovae at their ends of the 
evolution \citep{hege03}. 

\begin{figure*}[htbp]
\begin{center}
\FigureFile(160mm,160mm){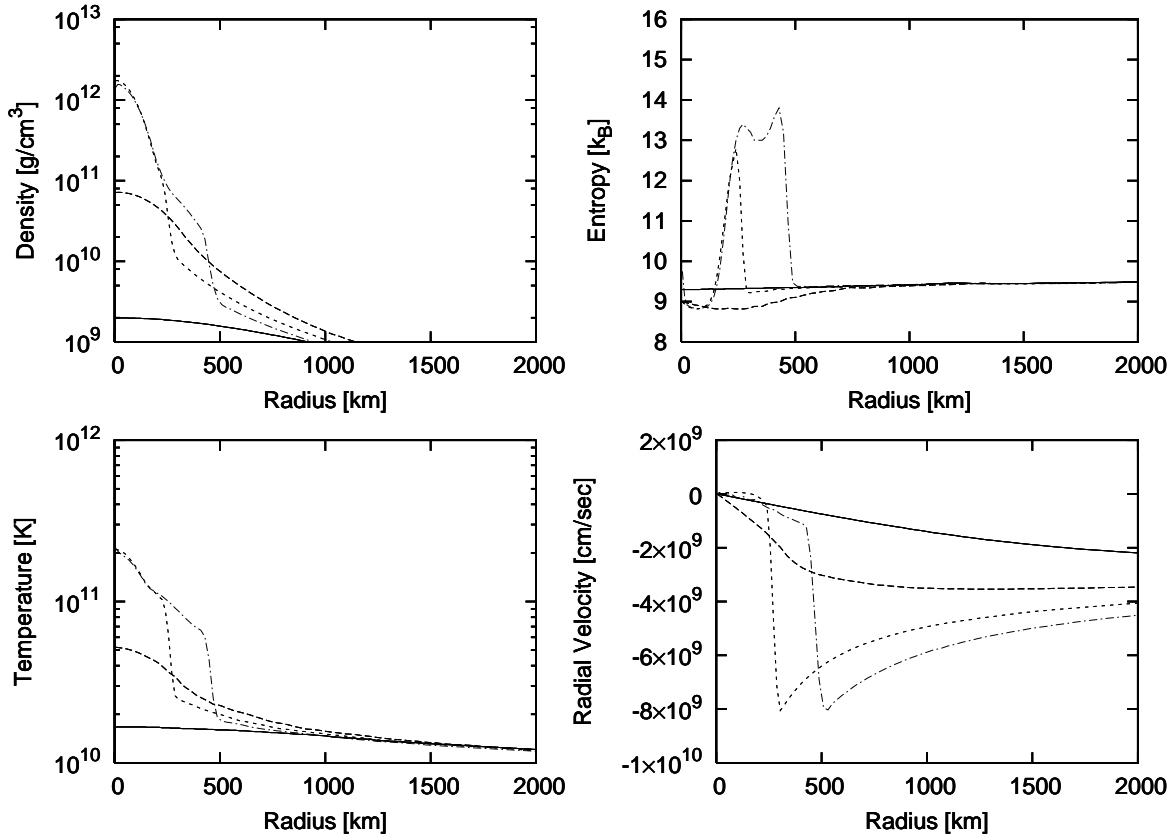}
\end{center}
  \caption{The evolutions of density, temperature, entropy 
and radial velocities in the spherical model.
  Solid line is for -37 ms from bounce, dashed line is for -1 ms, dotted line is for 19 ms and dashed dotted line is for 32 ms, respectively.}
 \label{fig:vel}
\end{figure*}

As in the case of massive stars, the gravitational collapse is triggered by the electron-capture reactions and the photodisintegration of iron nuclei.
On the other hand, the gravitational contraction is stopped not by the nuclear forces as in the case of massive stars but by the (gradient of) thermal pressure. This is because the progenitor of Pop III stars has high entropy, i.e. high temperature.
We call this bounce as ``thermal bounce'' for convenience.
The evolution of density, temperature, entropy and radial velocity around 
the thermal bounce are shown in Figure \ref{fig:vel}.
Unlike the case of massive stars, no outgoing shock propagates outward 
after the thermal bounce. At the bounce, the size of the inner core, 
which is 200 km in radius and $6M_\odot$ in the masscoordinate, 
grows gradually due to the mass accretion. 
As seen from the figure, the materials in the accreting shock regions obtain higher entropy and temperature than the ones in the case of massive stars.

\begin{figure}[htbp]
\begin{center}
\FigureFile(80mm,80mm){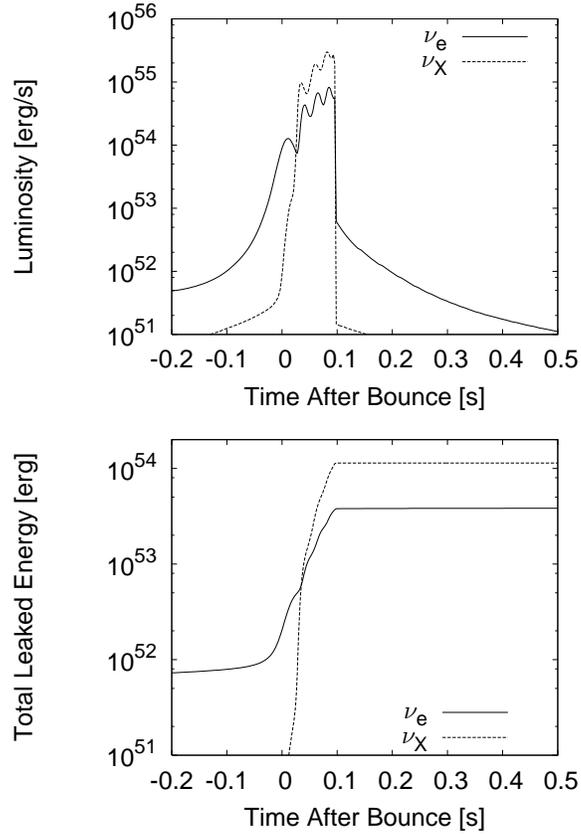}
\end{center}
  \caption{{\itshape Upper panel}: Time evolutions of neutrino luminosity in the spherical model.
  The time is measured from the thermal bounce.
  Solid line represents the total luminosity of electron neutrinos and anti-electron 
neutrinos.
  Dashed line represents the total luminosity of $\nu_X$ ($\nu_\mu$, $\nu_\tau$, $\bar\nu_\mu$ and $\bar\nu_\tau$) neutrino.
  Before the thermal bounce, the luminosity of electron + anti-electron neutrinos dominate that of $\nu_{X}$ luminosity, though, after the bounce, it reverses.  
  At $\sim$ 0.1 second after the bounce, the luminosities drastically decrease 
due to the BH formation.
  {\itshape Lower panel}: Time-integrated neutrino luminosities.
  Solid line and Dashed line are total energy emitted by electron neutrinos 
  and anti-electron  neutrinos and X neutrinos, respectively.}
  \label{fig:neut}
\end{figure}

The higher temperatures are good for producing a large amount of 
$\mu$- and $\tau$- neutrinos through the pair annihilation of electrons 
and positrons.
This also makes different features of the neutrino emissions from the case of 
massive stars, in which the electron-neutrino ($\nu_e$) 
luminosity dominates over those of the other 
species near the epoch of core bounce. 
As shown in the top panel of Figure \ref{fig:neut}, 
the total luminosity of $\nu_X$ ($\nu_\mu$, $\nu_\tau$, $\bar\nu_\mu$ and 
$\bar{\nu}_\tau$) begins to dominate over the total luminosity of 
electron neutrinos and anti-electron neutrinos at 25 msec after bounce 
(see the first intersection of the lines in the figure).  

At 87 msec after the bounce, the core is so heavy that it promptly 
collapses to a BH. 
\begin{figure}[htbp]
\begin{center}
\FigureFile(80mm,80mm){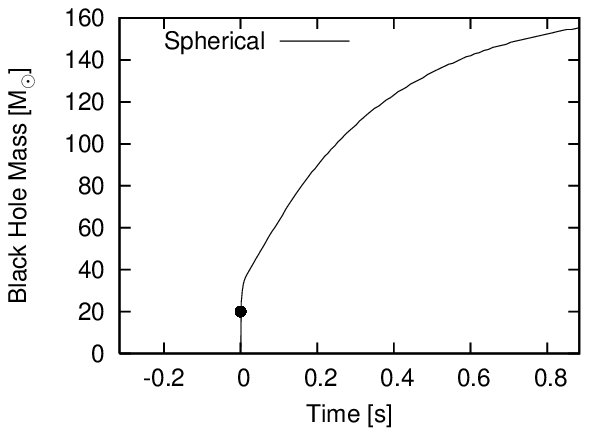}
\end{center}
  \caption{Evolution of the BH mass in the spherical model.
  The time is measured from the BH formation.
  The black filled circle indicates the epoch of the black-hole formation.}
  \label{fig:bh_sp}
\end{figure}
In Figure \ref{fig:bh_sp}, we show the evolution of the BH
 calculated by the procedure described in \S2.2.
The mass of the BH is initially 20$M_\odot$, rapidly increases to 
35 $M_\odot$ because the rest of the dense inner core falls into the BH 
soon after the formation. The growth rate of the mass is slowed 
down afterwards when the quasi-steady accretion flow to the BH is
 established (Fig \ref{fig:bh_sp}).
The rapid decrease of the neutrino luminosity $\sim 100$ msec after 
bounce (top panel of Figure \ref{fig:neut}) corresponds to the epoch 
when the neutrinospheres are swallowed into the BH.
Note in the bottom panel, the total 
emitted energy are calculated by $ \int \mathrm{dt} L_\nu$, 
which represents the energy carried out from the core by neutrinos.
Again from the quantity, it is shown that $\nu_X$ deprives dominantly of 
the gravitational energy of the core than $\nu_e$ and $\bar{\nu}_e$.

\subsection{Rotational Collapse}

\subsubsection{Effect of Differential Rotation}

Now we move on to discuss the features in the rotational core-collapse.
The deviation of the dynamics from the spherical collapse comes from 
the initial rotation rates and the degree of the differential rotation 
initially imposed.
To see the effects of the differential rotation on the collapse-dynamics, 
we first take models of B10TW1D (differential rotation) and B10TW1R 
(rigid rotation) as examples and mention the difference of them.
The effects of the initial rotation rates are discussed later in \S3.2.2.

We first describe the collapse of model B10TW1D.
As in the case of spherical collapse, the rotating core experiences the 
collapse due to the neutrino emission and the photodisintegration, but 
the difference appears at the time of the thermal bounce. 
Due to the pressure support supplied by the centrifugal force,
 model B10TW1D bounces at the pole at the epoch 17 msec later 
than that of the spherical collapse.
\begin{figure*}[htbp]
\begin{center}
\FigureFile(120mm,160mm){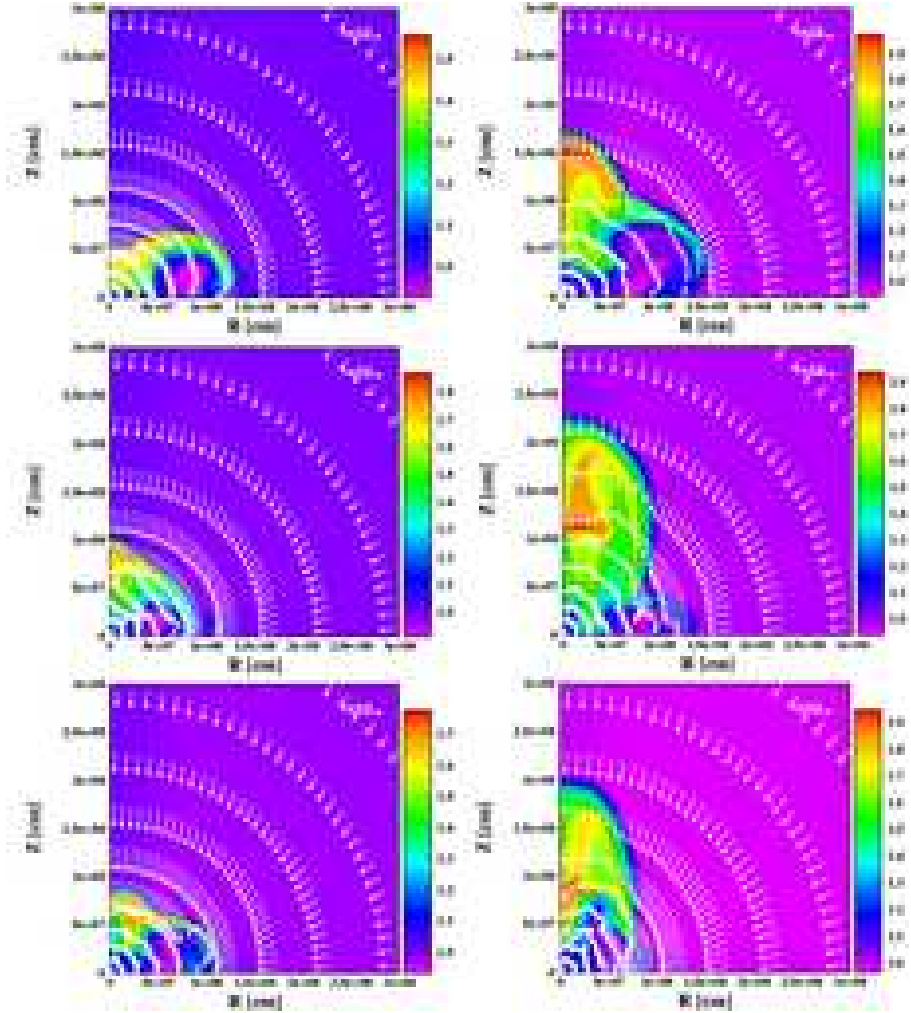}
\end{center}
  \caption{Entropy profiles of differential rotation model of B10TW1D 50 (left top), 63 (left middle), 73 (left bottom), 87 (right top), 113 (right middle), and 127 (right bottom) ms after bounce, respectively. The color coded contour shows the logarithm of entropy ($k_B$) per nucleon and arrows represent the velocity fields.}
  \label{fig:osci_ent}
\end{figure*}
The time evolutions after bounce is presented in Figure \ref{fig:osci_ent}.
It is shown that the materials of the inner core oscillate about 20 msec 
after bounce (see from the top left down to the bottom), and then
 the shock wave begins to propagate along the rotational axis (see from the 
top right down to the bottom).
This jet-like shock wave finally stalls at $Z \sim 2\times10^8$ cm, where 
$Z$ is the distance from the center along the rotational axis. 
It is noted that the shock wave formed at bounce 
does not stall in the strongly magnetized models as discussed in \S3.3.
In this weakly magnetized model, the stellar mantle just collapses to the 
central region after the shock-stall, and then leading to the formation of the BH.
In this model, we follow the hydrodynamics until more than 99 \% of the materials outside collapse to the BH (typically 2 sec after bounce).

\begin{figure*}[htbp]
\begin{center}
\FigureFile(120mm,120mm){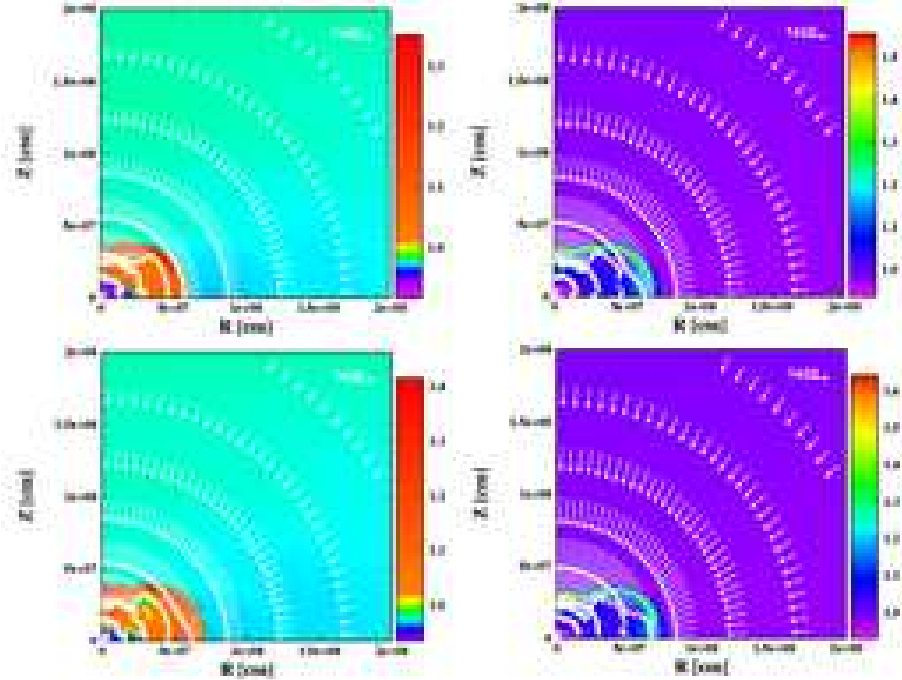}
\end{center}
  \caption{Same as Figure \ref{fig:osci_ent} but for the rigidly rotating model of 
B10TW1R at 
56 (left top), 76 (left bottom), 106 (right top), and 134 (right bottom) ms after bounce, respectively. Color coded contour shows the logarithm of entropy ($k_B$) 
per nucleon and arrows represent the velocity fields.}
  \label{fig:osci_ent_r}
\end{figure*}

Model B10TW1R thermally bounces rather isotropically in the center, not like 
model B10TW1D.
This is because the central regions have less angular momentum in comparison 
with the differentially rotating model of B10TW1D.
In Figure \ref{fig:osci_ent_r}, the time evolutions of entropy
 after bounce are shown.
Unlike B10TW1D (Figure \ref{fig:osci_ent}), B10TW1R directly 
collapses to form the BH without producing the outgoing shock waves.
This is because the central part has less pressure support from the 
centrifugal force due to the uniform rotation profile initially imposed.
On the other hand, the model B10TW1R has more angular momentum than 
that of model B10TW1D in the outer part of the core. This leads to the 
suppression of the accretion rates of the infalling matter to the inner core.
As a result, the core of model B10TW1R oscillates in a longer period than 
that of model B10TW1D because the dynamical timescale, which is proportional 
to $\rho^{-1/2}$, becomes longer due to the smaller density there.  

\begin{figure}[htbp]
\begin{center}
\FigureFile(80mm,80mm){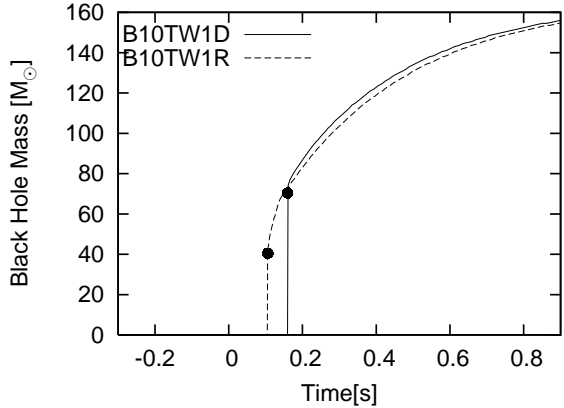}
\end{center}
  \caption{Evolution of the masses of BH for the rotating models.
  Solid and dashed lines are for models B10TW1D and B10TW1R, respectively.
  The black circle indicates the epoch of the black-hole formation.
  Note that the time is measured from the epoch of the BH formation in the 
spherical model. }
  \label{fig:bh_rot_dr}
\end{figure}

Next we compare the masses of the BH at the formation and 
the subsequent growth between the two models (see Figure 
\ref{fig:bh_rot_dr}).  The initial mass of BH of models B10TW1R and 
  B10TW1D are 40 and 70 $M_\odot$, respectively 
(see the black filled circles in the Figure). Both of them are larger 
than that of the spherical collapse model ($\sim 20~ M_\odot$). As mentioned, 
the 
reason that the earlier formation of less massive BH of model B10TW1R is 
 that the model has smaller centrifugal forces in the central regions than  
model B10TW1D. On the other hand, reflecting the smaller mass accretion rates 
 to the BH, 
the growth rate of BH's mass of the model B10TW1R is smaller than that of 
the model B10TW1D (compare the slopes of the lines in the Figure 
after the BH formation). 

\begin{figure*}[htbp]
\begin{center}
\FigureFile(160mm,160mm){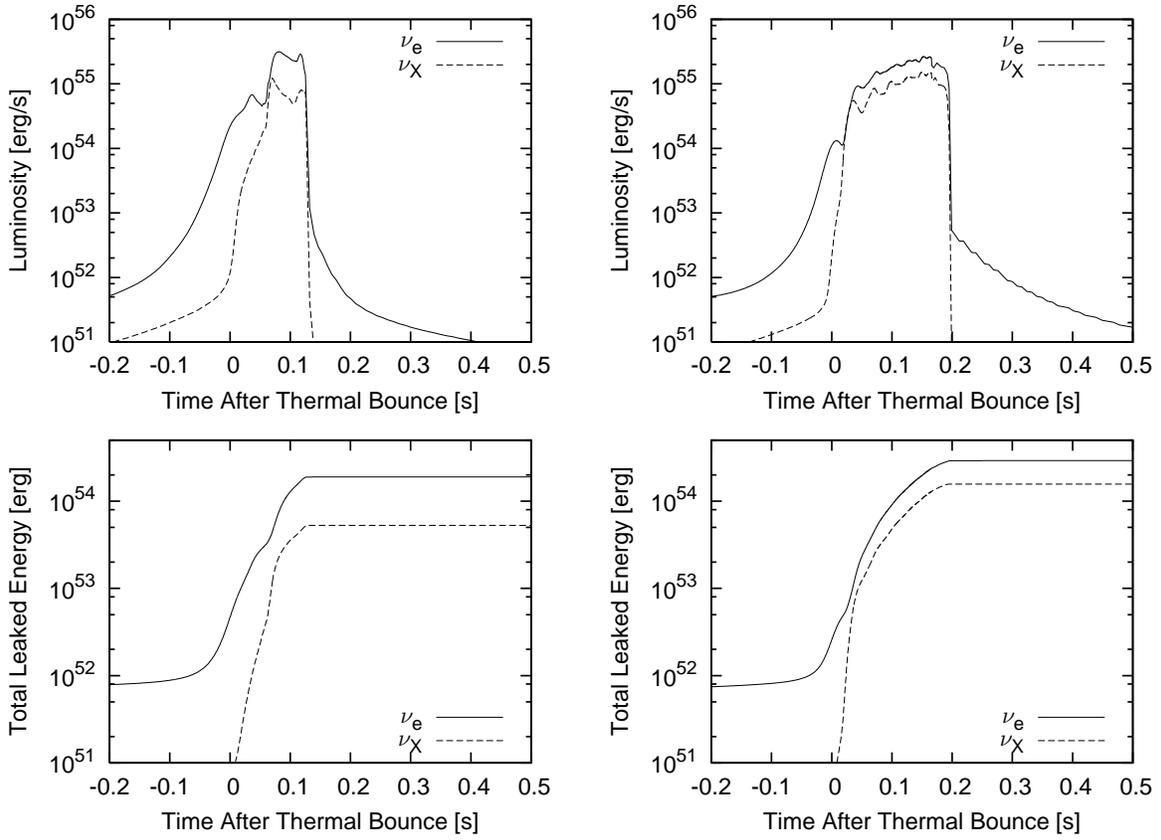}
\end{center}
  \caption{Same as Figure \ref{fig:neut} but for models 
B10TW1D (left) and B10TW1R (right), respectively.}
  \label{fig:neut_lum_rot}
\end{figure*}

The luminosity of neutrinos and the total leaked energy of the model B10TW1D 
are shown in the left panel of Figure \ref{fig:neut_lum_rot}.
It is found that the luminosity of $\mu$ and $\tau$ neutrinos ($\nu_X$) do not 
overwhelm that of the electron neutrinos even after the bounce unlike 
the spherical collapse and that most energy are emitted by electron neutrinos
 (bottom panel).
This is because the rotation suppresses the compression of the core, 
which lowers the temperature in the central regions than that of the spherical model. It should be noted that the energy production rates 
by the pair annihilation processes sharply depend on the temperature. 
The neutrino features of B10TW1R are found to be intermediate 
between the model B10TW1D and the spherical model (see right panels).

If the initial rotation rates of the above two models become larger, 
the bounce occurs more later due to the stronger centrifugal forces. 
In addition, the interval of the core oscillations becomes longer.
Except for such differences, the hydrodynamic features before the BH formation 
are mainly determined by the degree of the differential rotation as mentioned 
above and no qualitative changes are found as the initial rotation 
rates become larger.

\subsubsection{Effects of Rotation on the BH mass and Neutrino emission}
In this section, we proceed to describe how the initial rotation rate and 
the degree of the differential rotation affect the growth of the BH masses 
and the neutrino emissions.

\begin{figure}[htbp]
\begin{center}
\FigureFile(80mm,80mm){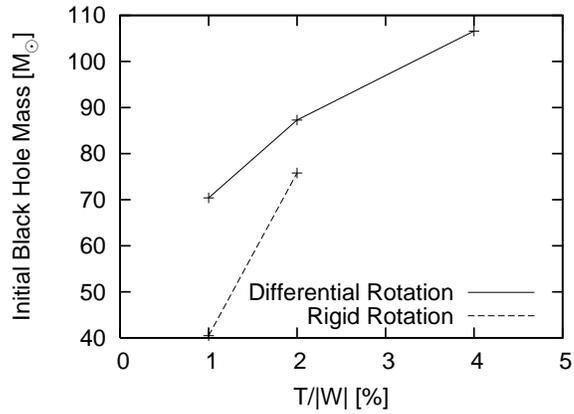}
\end{center}
  \caption{Effects of initial rotation rate and the degree of differential rotation on the initial mass of the BH. In this figure, the sequence of the models
 labeled by ``B10'', which are almost purely rotating model, is chosen.
 Note that model B10TW4R is absent because this model does not
produce the BH during the simulation time.}
  \label{fig:BH_rot_tw}
\end{figure}
 The effects of rotation on the initial masses of the BHs for the 
almost purely rotating models, labeled by B10, are shown in Figure 
\ref{fig:BH_rot_tw}.
As seen, larger the initial rotation rate becomes, the heavier BH is 
found to be produced. This tendency is independent of  
the degree of the differential rotation. This is simply because rapid 
rotation tends to halt the infall of the matter to the center, thus heavier 
masses are required to fulfill the condition of BH formation. 
It is furthermore found that the initial mass is larger 
for the differential rotation models than the rigid rotation models. 
This is regardless of the initial rotation rates. 
This is due to the smaller angular momentum of the rigid rotation models 
in the central regions than that of the differential rotation models 
as mentioned.

\begin{figure*}[htbp]
\begin{center}
\FigureFile(160mm,80mm){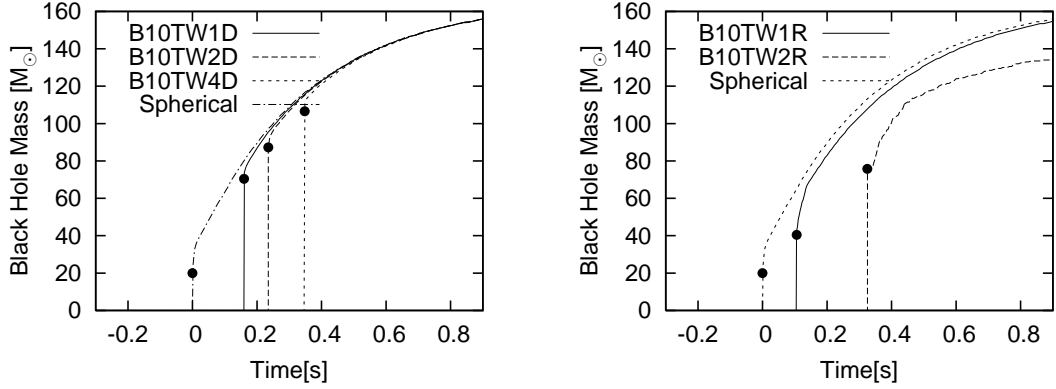}
\end{center}
  \caption{Time evolution of the BH mass for the almost purely rotating models 
labeled by B10.
    {\itshape Left panel}:
    Solid, dashed, dotted, and dashed-dotted lines, are for models B10TW1D, B10TW2D, B10TW4D, and the spherical model, respectively.
    {\itshape Right panel}:
    Solid, dashed, dotted lines are for models B10TW1R, B10TW2R, and the spherical model, respectively.
  The time is measured from the thermal bounce of each model.
  The black filled circles of each panel represent the epoch of the BH formation.
}
  \label{fig:BH_rot_dr-2}
\end{figure*}

In Figure \ref{fig:BH_rot_dr-2}, the growth of the BH mass 
for the corresponding models is shown.
It is found that the epoch of the formation is delayed as 
the initial rotation rates become larger regardless of the degree of the 
differential rotation. As for the growth rates of BH's mass,
 it is found that they are almost the same for the differential rotation 
models regardless of the initial rotation rates 
(see the left panel of Figure \ref{fig:BH_rot_dr-2}). 
This is because the outer 
part of the core has little angular momentum due to the strong differential 
rotation imposed, and thus falls to the center in the similar way.
On the other hand, the initial rotation rates affect the 
evolution of BHs in the rigid rotation models (see the right panel of Figure 
\ref{fig:BH_rot_dr-2}). As the initial rotation rates become larger,
 the growth rates of the BHs become smaller due to the larger angular momentum 
imposed initially.

\begin{figure*}[htbp]
\begin{center}
\FigureFile(160mm,80mm){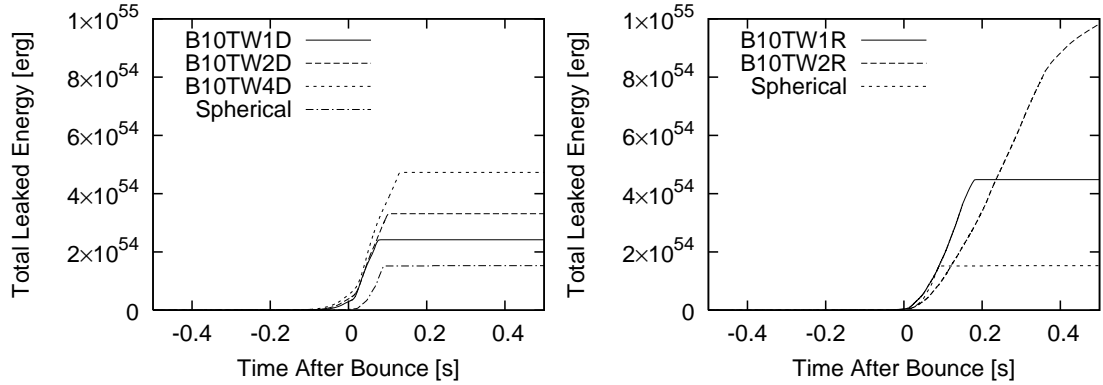}
\end{center}
  \caption{Time-integrated neutrino leakage energy.
    {\itshape Left panel}:
  Solid line means B10TW1D, dashed line means B10TW2D, dotted line means B10TW4D  and dash-dotted line means Spherical model.
    {\itshape Right panel}:
  Solid line means B10TW1R, dashed line means B10TW2R and dotted line means Spherical model.
  The time is measured from the thermal bounce of each model.
  }
  \label{fig:neut_rot}
\end{figure*}

BH's mass at the formation affects the total energy emitted by the 
neutrinos because the neutrinos in the region of BH cannot escape to the 
outside of the core afterwards.
The total energy emitted by neutrinos are shown in Figure \ref{fig:neut_rot}. One can see the general trend in the figure that the 
 emitted energy rapidly rises and then becomes constant.
The transition to the constant phase corresponds to the formation of the BH.

Also in this case, differential rotation models have similar features 
after the formation of the BHs 
 (solid line of right panel of Figure \ref{fig:neut_rot}). 
It is interesting that the model B10TW2R 
(dashed line of right panel of Figure \ref{fig:neut_rot}) by contrast has 
different behaviors of total emitted energy. This is because 
this model produces the stable accretion disk around the central BH. As 
a result, the materials of the disk accretes only slowly to the BH, and 
thus can emit neutrinos for a longer time.
The small accretion rate of this model is also prominent as seen in 
the Figure \ref{fig:BH_rot_dr-2}.

It is interesting to note that only about 10\% of the gravitational energy of 
the core can be carried away by neutrinos even in the most rapidly 
rotating model 
considered here (B10TW4D). On the other hand, it is well known that 
neutrinos carry away $99~ \%$ of the gravitational energy of the protoneutron 
stars in case of the massive stars. The discrepancy stems obviously from the 
fact that most part of the inner core is absorbed to the BH in case of the 
Pop III stars.

\subsection{Magnetorotational Collapse}
In this section, we present the results of the MHD models. 
First of all, we mention the magnetohydrodynamic (MHD) features
 in section \ref{mhd}, then discuss the MHD effects on the BH mass and neutrino  emissions in section \ref{MEB}. 

\subsubsection{MHD Feature \label{mhd}}
Amongst the computed models, we find that the models with the strongest magnetic field 
($B=10^{12}$G) can only produce the jet-like shock waves
along the rotational axis, which can propagate outside of the core without 
shock-stall. 
First of all, we mention the properties of such models 
taking model B12TW1D as an example. 

The collapse dynamics before bounce is almost the 
same as the corresponding weak magnetic field model of B10TW1D. 
This is because the amplified magnetic fields by the compression and 
the field-wrapping are, of course larger than the weaker field model, 
but still much smaller than the matter pressure in the central regions. 
After the bounce, the toroidal magnetic fields produced by the wrapping, provide the additional pressure support,  thus acting to push the infalling matter as jetlike outflow rather than rotates along the magnetic field. The jet is launched when the magnetic pressure overcomes the local ram pressure of the accreting matter. This feature is different from another jet driving mechanism, the magneto-centrifugal acceleration \citep{blan82}.

\begin{figure*}[t]
\begin{center}
\FigureFile(120mm,160mm){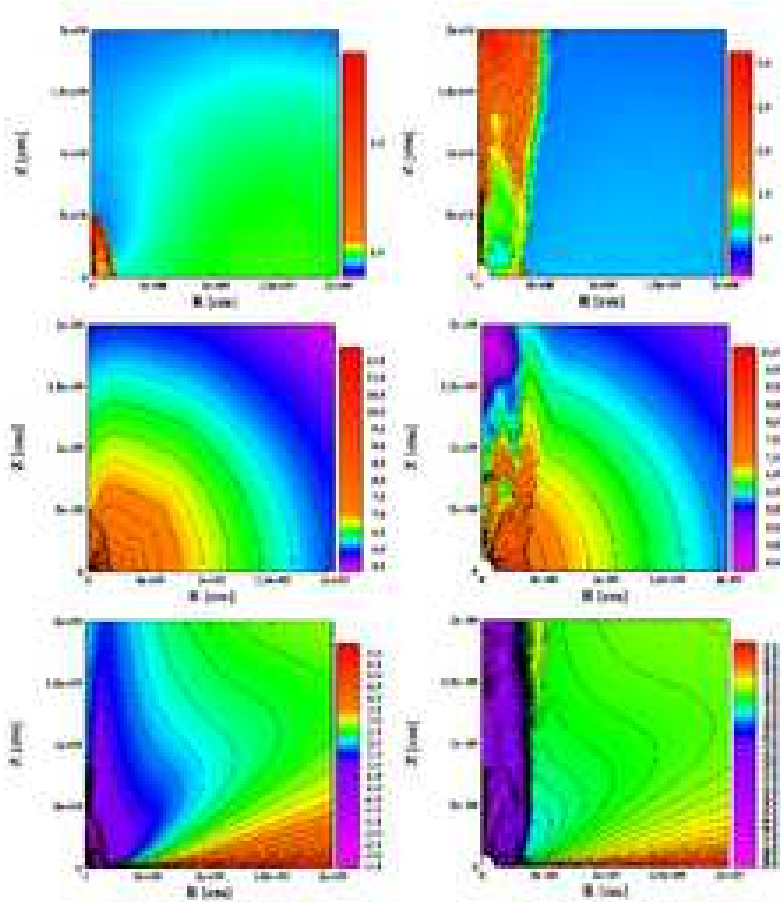}
\end{center}
  \caption{Time evolution of shock waves of the strongest magnetized model of B12TW1D. The top panel of the figure shows the logarithm of entropy ($k_B$) per nucleon, the middle panel of the figure shows logarithm of density (g cm$^{-3}$), and bottom panel shows the logarithm of plasma beta. All left figures are at 119 ms from bounce, and the right figures are at 305 ms. The white circles of the right panels represent the BHs.}
  \label{fig:B12TW1D}
\end{figure*}

The MHD features of model B12TW1D after bounce 
are presented in Figure \ref{fig:B12TW1D}.
From the right panels, it is shown that the regions behind the 
jet-like shock wave ($Z \geq 1.5 \times 10^{9}~{\rm cm} $) become 
dilute with the density of $\rho \sim 10^{5} ~{\rm g}~{\rm cm}^{-3}$ 
and have very high entropy $s \sim 10^{2} k_B$.
The bottom panel shows the jet is driven by magnetic pressure because the 
plasma beta ($\equiv$ gas pressure / magnetic pressure) of the region of 
jet is much smaller than unity.
As the jet propagates in the core, a newborn BH is produced 
(see the white circles of right panels of Figure \ref{fig:B12TW1D}).
The mass of the BH is initially 57.9 $M_\odot$, which is smaller than the 
one of B10TW1D (70.4$M_\odot$).
The reason of the difference is mentioned in \S\ref{MEB}.

\begin{figure*}[t]
\begin{center}
\FigureFile(160mm,120mm){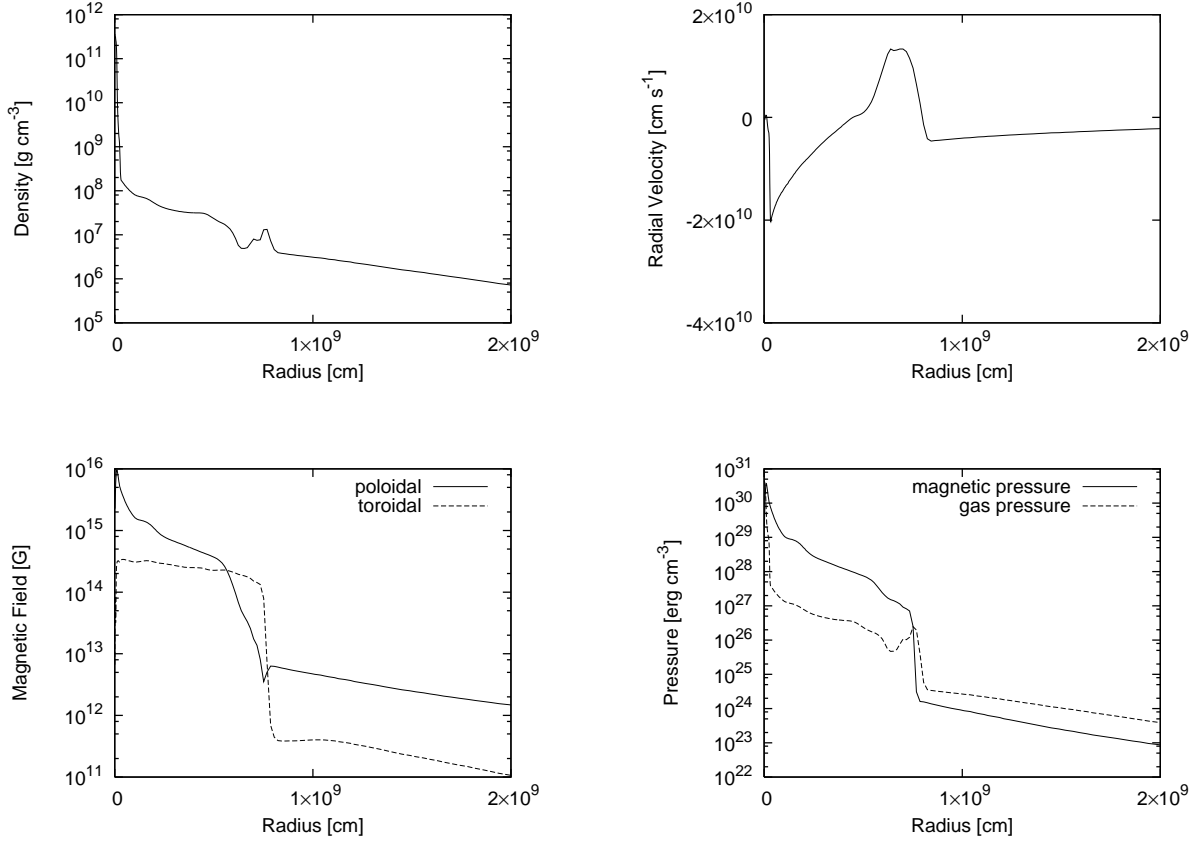}
\end{center}
  \caption{
  Various physical quantities around the rotational axis at 104 ms 
  after bounce for model B12TW1D. 
  Density (left top), radial velocity (right top), absolute value of 
  magnetic field (left bottom), and pressure are shown.
  In the left bottom panel, the solid line and dashed line represent 
  poloidal component and toroidal component, respectively.
  In the right bottom panel, the dashed line represents gas pressure 
  and solid line represents magnetic pressure.
  }
  \label{fig:jet_prof}
\end{figure*}

The properties of jet of model B12TW1D are shown in Figure 
\ref{fig:jet_prof}. There are profiles of density, radial velocity, 
magnetic field, and pressure at 104 ms after bounce. 
The density of matter in the jet region is $\sim 10^7$ g cm$^{-3}$.
The speed of shock front is as large as 40 \% of speed of light, 
which is mildly relativistic. 
It is easily seen that the toroidal magnetic field overwhelms the 
poloidal component behind the shock front.
In the inner region, the poloidal magnetic field is larger due to 
compression.
The magnetic pressure overwhelms gas pressure throughout the jet 
region as already depicted in Figure \ref{fig:B12TW1D}.

\begin{figure*}[htbp]
\begin{center}
\FigureFile(120mm,120mm){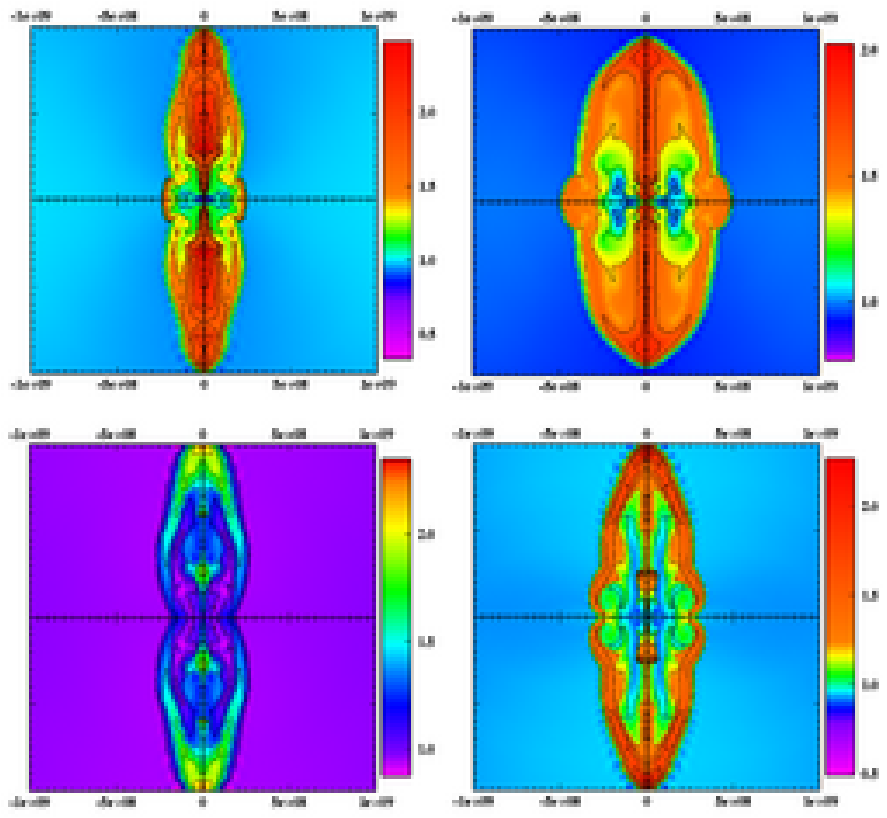}
\end{center}
  \caption{Profiles of the shock propagation for models B12TW1D (top left), B12TW4D (top right), B12TW1R (bottom left), and B12TW4R (bottom right), respectively. They show the color coded contour plots of logarithm of entropy ($k_B$) per nucleon. Various profiles are found by changing the strength of the initial magnetic field and rotation.}  
  \label{fig:B12C}
\end{figure*}

Now we move on to discuss how rotation affects the dynamics while fixing 
the initial field strength.
Figure \ref{fig:B12C} shows the properties of models B12TW1\{D,R\} and 
B12TW4\{D,R\} when the jet-like shock wave reaches to $1\times 10^{9}$ cm.
As clearly seen, the main difference between 
$T/|W|$ = 1 \% and 4 \% is the degree of the collimation of the shock wave.
As the initial rotation rates become large
 in the differential rotation models, 
the compression of the magnetic fields is hindered, thus leading to the 
 suppression of the hoop stress of magnetic fields in the central regions 
\citep{taki04}. As a result, the collimation of the shock wave becomes less 
(compare the top panels). 
 In contrast, the difference of the degree of collimation between 
models B12TW1R and B12TW4R is smaller than B12TW1D and B12TW4D. This is 
because the materials in the inner region of rigid rotating models rotate 
more slowly than the differential models. Thus 
the degree of the collimation of the shocks depends weekly on the initial 
$T/|W|$.

Models with the weaker initial magnetic fields do not produce the 
jet-like explosion except for model B11TW4R.
These models collapse to BHs before formation of jets because of weak 
magnetic pressure. After forming a BH, especially differentially rotating 
model, rest parts of star rotate slowly so that the magnetic pressure does 
not grow up.
Thus, when BH is formed before jet rises, 
rest of core only collapse to BH and entire the core are absorbed by BH.

\subsubsection{MHD Effects on the BH Mass and Neutrino Emission \label{MEB}}

\begin{figure*}[t]
\begin{center}
\FigureFile(80mm,80mm){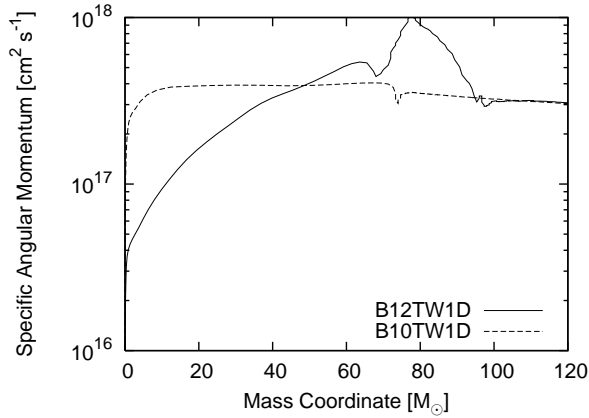}
\end{center}
  \caption{ Mean specific angular momentum over the shells as a function of the mass coordinate
   just before BH formation.
  The solid line and dotted line represent model B12TW1D and  B10TW1D, respectively.}
  \label{fig:ang_mom}
\end{figure*}

\begin{figure*}[htbp]
\begin{center}
\FigureFile(80mm,80mm){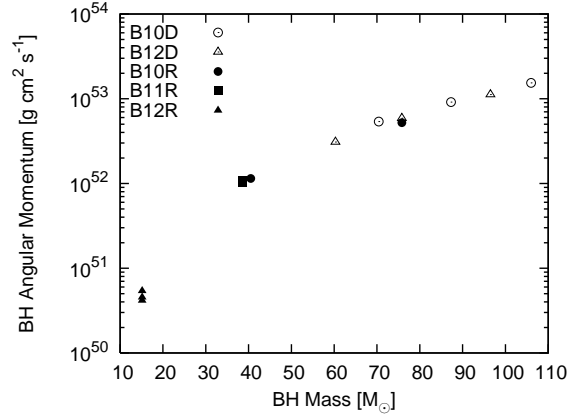}
\end{center}
  \caption{Relation between BH mass and angular momentum at the BH formation.}
  \label{fig:mass_am}
\end{figure*}

\begin{table}[htbp]
\begin{center}
\caption{Initial Mass of Black Holes for Differentially Rotating Models [$M_\odot$].}
\label{tab:bh1}
\begin{tabular}{|c|ccc|}
\hline\hline
 &\multicolumn{3}{c|}{$T/|W|$} \\
\cline{2-4}
$B_{0}$ & 1\% &2\%  &4\%\\
\hline
$10^{10}$G  & 70.4  & 87.3 & 106.6  \\
$10^{11}$G  & 70.4  & 87.3 & 106.6  \\
$10^{12}$G  & 57.9  & 75.8 & 96.6   \\
\hline
\end{tabular}
\caption{Initial Mass of Black Holes for Rigidly Rotating Models [$M_\odot$].}
\label{tab:bh2}
\begin{tabular}{|c|ccc|}
\hline\hline
 &\multicolumn{3}{c|}{$T/|W|$} \\
\cline{2-4}
$B_{0}$ & 1\% &2\%  &4\%\\
\hline
$10^{10}$G  & 40.5  & 75.8 & ---  \\
$10^{11}$G  & 38.6  & 38.6 & ---  \\
$10^{12}$G  & 15.3  & 15.1 & 15.1  \\
\hline
\end{tabular}
\end{center}
\end{table}
The MHD effects on the initial masses of BHs are shown in Tables \ref{tab:bh1} 
and \ref{tab:bh2}.
As seen, the initial BH mass gets smaller when the initial magnetic 
field becomes stronger.
The angular momentum transport by the magnetic fields 
is an important agent to affect the BH mass.
This feature is seen in Figure \ref{fig:ang_mom}, which represents 
the distribution of mean specific angular momentum of models B12TW1D 
and B10TW1D. The central region of B12TW1D has smaller angular 
momentum than B10TW1D due to angular momentum transport by magnetic 
fields. The peak of model B12TW1D represents the position of shock 
front on the equatorial plane.
The transport of the angular momentum makes centrifugal force of 
central region smaller and enhances the collapse.
This leads the BH mass smaller.
This tendency is more prominent for the rigid rotation models 
(compare Tables \ref{tab:bh1} and \ref{tab:bh2}) because the rotation 
of the central region is slower than differential rotation models and 
contraction of core is more significant, which leads the 
amplification of magnetic field and larger angular momentum transport.
Figure \ref{fig:mass_am} shows the relation between BH mass 
and angular momentum at BH formation. 
The angular momentum of BH gets larger with its mass at the time of BH formation.
This is because the matter with large angular momentum cannot collapse 
due to centrifugal force and requires large amount of accreting matter 
to collapse BH, as already mentioned in \S 3.2.

\begin{figure}[htbp]
\begin{center}
\FigureFile(80mm,80mm){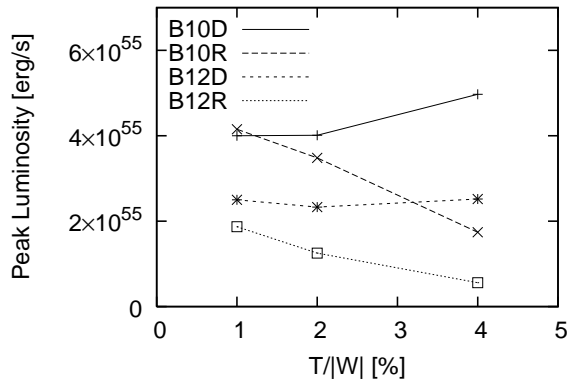}
\end{center}
  \caption{Effects of rotation and magnetic fields on the peak luminosity of neutrinos.}
  \label{fig:peak_lumi}
\end{figure}

Next, we discuss the MHD effects of neutrino emissions.
Figure \ref{fig:peak_lumi} shows the peak neutrino luminosities as 
a function of the initial rotation rates. It is shown that the magnetic 
fields make the peak luminosities smaller, when fixing the initial degree
 of the differential rotation (compare B10D and B12D, and B10R and B12R).
 It is noted that the Pop III stars have gentle slope of the 
density prior to core-collapse, so that materials of the 
outer region have a great deal of the total gravitational energy 
of the iron core. 
Thus the stronger magnetic pressures, which  
 prevent the accretion, make the liberating 
gravitational energy of the accreting matter smaller, and thus, 
results in the suppression of the peak luminosities. In case of the 
rigid rotation, the stronger centrifugal forces in the outer regions, lead 
to the stronger suppression of the releasable gravitational energy than in the 
case of the differential rotation. As a result,
 the peak luminosities
 for the rigidly rotating models decreases more steeply with the initial 
rotation rates than the ones for the differentially rotating models (compare B10R and B10D).

%

\section{SUMMARY AND DISCUSSION}

We studied the magnetorotational core-collapse of Pop III stars 
by performing the two-dimensional magnetohydrodynamic simulations.
Since the distributions of rotation and magnetic fields in the progenitors of Pop III stars are highly uncertain, we changed them in a parametric manner and systematically investigated how rotation and magnetic fields affect the dynamics of Pop III stars.
In addition, we explored how rotation and magnetic fields affect the formation of the BHs and 
the neutrino emissions.
In the current Newtonian simulations, the BH formation was ascribed to a certain condition, and after the formation, the central region was excised and treated as an absorbing boundary.
As for the microphysics, we took into account the neutrino cooling of 6 species by a leakage scheme with a realistic equation of state.
With these computations, we have obtained the following results,

\begin{enumerate}
\item
In the spherical model, the gravitational contraction is stopped by the gradient of the thermal pressure, not by the nuclear forces as in the case of the massive stars $\approx O(10) M_{\odot}$ with the initial composition of the solar metallicity, because the progenitor of Pop III stars has high 
entropy, i.e. high temperature initially.
Such high temperature also makes different features of the neutrino emissions from the case of the massive stars.  
The luminosity of $\mu$- and $\tau$- neutrinos dominates over that of the electron neutrinos after core bounce.
Thus the gravitational energy of the core is carried away dominantly by  $\mu$- and $\tau$- neutrinos.

\item
As the initial rotation rates of the core become larger, it is found that the epoch of the BH formation is delayed later and that the initial masses of the BHs become larger.
Fixing the initial rotational energy, the BH masses at the formation become larger as the degree of the differential rotation becomes stronger.
As the initial degree of the differential rotation becomes larger, the electron neutrino luminosity is found to be more dominant over that of $\mu$ and $\tau$ neutrinos after core bounce, because the
 pair creation processes of $\mu$ and $\tau$, which sharply depend on the 
temperature, are more suppressed.

\item
We find that the jet-like explosions can be produced even in Pop III stars if the magnetic field is as large as $10^{12}$G prior to core-collapse.
This jet-like shock wave is completely magneto-driven.

\item
Jet-like shocks in the stronger magnetic field models are themselves found to make the initial mass of the BH smaller.
The angular momentum transport by magnetic fields is found to be an important agent to make the initial mass of BHs smaller because the transport 
of the angular momentum enhances the collapse of the central regions.
As a result, it is found that the initial BH masses for the most strongly magnetized models are found to become smallest when fixing the initial rotation rates. 
As for the neutrino luminosities, we point out that the stronger magnetic fields make the peak luminosities smaller, because they can halt the collapse of the materials.
\end{enumerate}

Here we shall refer to the limitations of this study. First of all, we mimicked the neutrino transfer by the leakage scheme. Although the scheme is 
a radical simplification, we checked that we could reproduce, at least, the qualitative features of the neutrino luminosities.
The supremacy of $\nu_X$ neutrinos' luminosity in the spherical collapse 
of Pop III stars obtained in this study is consistent with the foregoing 
studies by \citet{fryer01} and \citet{naka06}, in which the more elaborate 
neutrino transport schemes were employed.
Furthermore, the effects of rotation on the emergent neutrino luminosities 
are consistent with \citet{frye00}, in which one model of the rotational 
collapse of the massive stars was investigated.
Secondly, the simulations were done with the Newtonian approximation and we 
defined the BH formation by the marginally stable orbit of a Schwarzschild BH.
This treatment is totally inaccurate because the core rotates so rapidly that 
fully general relativistic (MHD) simulations with the appropriate 
implementations of the microphysics are necessary, 
however, are still too computationally prohibitive and 
beyond our scope of this paper.
Remembering these caveats, this calculation is nothing but a demonstration 
showing how the combinations of rotation 
and magnetic fields could produce the variety of the dynamics, and the 
important consequences in the properties of the neutrino emissions, and 
these outcomes, of course, should be re-examined by the more sophisticated 
simulations.

In this study, we followed the dynamics till $\sim 1$ sec after the formation 
of the BHs and saw the shock break-out from the cores
 in the strongly magnetized models. But if we were to follow the dynamics 
in the much later phase in the weaker 
magnetized models, the magneto-driven outflows might be produced, due to the 
 long-term field-wrapping and/or the development of the 
so-called magnetorotational instability, as 
 demonstrated in the study of collapsar (see, e.g., \cite{prog03b,
fuji06}). 
 But it should be noted that the dynamical phases considered here 
 and the other ones are apparently different (and thus they are complimentary).
In the latter studies, the central BHs with a rotationally 
supported disk around are treated as an initial condition for the computations. Our core-collapse simulations presented here showed that the outer region 
rotates much slowly than 
the Keplerian one and most of them directly collapses to BH. Thus   
the amplification of the magnetic field in the disks, which needs 
rapid rotation, might not be so efficient as previously demonstrated. To clarify it, we are now preparing for the long-term 
simulations, in which the final states obtained here are taken as an 
initial condition. 

Then we discuss the validity of the initial strength of the magnetic fields 
assumed in this study.
For the purpose, we estimate the strength of the magnetic field just before collapse with Eq. (13) of \citet{maki04}, in which the thermal history of the primordial collapsing clouds was calculated in order to investigate the coupling of the magnetic field with the primordial gas.
For example, $B_\mathrm{ini}\sim10^{-7}$ G and $n_{\mathrm{H,ini}}\sim10^3~ \mathrm{cm}^{-3}$, which are the values they employed, lead $B\sim10^{11}$ G if the magnetic flux is conserved during the contraction and the clouds collapse to $10^{6}~ \mathrm{g~cm}^{-3}$.
Although the above parameters chosen are slightly optimistic, the magnetic fields assumed in this study may not be so unrealistic.

We pointed out that the total neutrino energy emitted from rotation models increases several times than the one of the spherical collapse model.
However, the detection of such neutrinos as the diffusive backgrounds might be difficult because the Pop III stars are too distant (see \cite{iocco05}). 
Alternatively, the detection of gravitational waves from Pop III stars as the backgrounds seems more likely \citep{buon05,sand06} by the currently planning air-borne laser interferometers such as LISA\footnote{http://lisa.jpl.nasa.gov}, DECIGO \citep{seto01} and BBO \citep{unga05}, and needs further investigation. 

We found that the Pop III stars are able to produce jet-like explosions with mass ejections when the central cores are strongly magnetized.
This may be important with respect to its relevance to the nucleosynthesis in such objects \citep{ohku06}. This is also an interesting topic to be 
investigated as a sequel of this paper.

\bigskip
 This study was supported in part by the Japan Society for 
Promotion of Science (JSPS) Research Fellowships (T.T.), 
Grants-in-Aid for the Scientific Research from the Ministry of Education, Science and Culture 
of Japan (No.S14102004, No.14079202, No.17540267, and No. 1840044).
Numerical computations were in part carried out on VPP5000 at the Center 
for Computational Astrophysics, CfCA, of the National Astronomical 
Observatory of Japan.


\begin{thebibliography}{47}
\expandafter\ifx\csname natexlab\endcsname\relax\def\natexlab#1{#1}\fi

\bibitem[{Abel {et~al.}(2002)Abel, Bryan, \& Norman}]{abel02}
Abel, T., Bryan, G.~L., \& Norman, M.~L. 2002, Science, 295, 93

\bibitem[{Akiyama {et~al.}(2003)Akiyama, Wheeler, Meier, \&
  Lichtenstadt}]{aki03}
Akiyama, S., Wheeler, J.~C., Meier, D.~L., \& Lichtenstadt, I. 2003, \apj, 584,
  954

\bibitem[{{Ando} \& {Sato}(2004)}]{ando04}
{Ando}, S. \& {Sato}, K. 2004, New Journal of Physics, 6, 170

\bibitem[{{Ardeljan} {et~al.}(2005){Ardeljan}, {Bisnovatyi-Kogan}, \&
  {Moiseenko}}]{arde05}
{Ardeljan}, N.~V., {Bisnovatyi-Kogan}, G.~S., \& {Moiseenko}, S.~G. 2005,
  \mnras, 359, 333

\bibitem[{{Barkana} \& {Loeb}(2001)}]{bark01}
{Barkana}, R. \& {Loeb}, A. 2001, \physrep, 349, 125

\bibitem[Blandford \& Payne(1982)]{blan82} Blandford, R.~D., 
\& Payne, D.~G.\ 1982, \mnras, 199, 883 

\bibitem[{Bond {et~al.}(1984)Bond, Arnett, \& Carr}]{bond84}
Bond, J.~R., Arnett, W.~D., \& Carr, B.~J. 1984, \apj, 280, 825

\bibitem[{Bromm {et~al.}(2002)Bromm, Coppi, \& Larson}]{brom02b}
Bromm, V., Coppi, P.~S., \& Larson, R.~B. 2002, \apj, 564, 23

\bibitem[{{Bromm} \& {Larson}(2004)}]{brom04}
{Bromm}, V. \& {Larson}, R.~B. 2004, \araa, 42, 79

\bibitem[{{Bromm} \& {Loeb}(2006)}]{brom06}
{Bromm}, V. \& {Loeb}, A. 2006, \apj, 642, 382

\bibitem[{{Buonanno} {et~al.}(2005){Buonanno}, {Sigl}, {Raffelt}, {Janka}, \&
  {M{\"u}ller}}]{buon05}
{Buonanno}, A., {Sigl}, G., {Raffelt}, G.~G., {Janka}, H.-T., \& {M{\"u}ller},
  E. 2005, \prd, 72, 084001

\bibitem[{{Christlieb} {et~al.}(2002){Christlieb}, {Bessell}, {Beers},
  {Gustafsson}, {Korn}, {Barklem}, {Karlsson}, {Mizuno-Wiedner}, \&
  {Rossi}}]{chri02}
{Christlieb}, N., et al.
  2002, \nat, 419, 904

\bibitem[{Daigne {et~al.}(2004)Daigne, Olive, Vangioni, Silk, \&
  Audouze}]{daig04}
Daigne, F., Olive, K.~A., Vangioni, E., Silk, J., \& Audouze, J. 2004, \apj,
  617, 693

\bibitem[{{Frebel} {et~al.}(2005){Frebel}, {Aoki}, {Christlieb}, {Ando},
  {Asplund}, {Barklem}, {Beers}, {Eriksson}, {Fechner}, {Fujimoto}, {Honda},
  {Kajino}, {Minezaki}, {Nomoto}, {Norris}, {Ryan}, {Takada-Hidai},
  {Tsangarides}, \& {Yoshii}}]{freb05}
{Frebel}, A., et al.
  2005, \nat, 434, 871

\bibitem[{{Fryer} \& {Heger}(2000)}]{frye00}
{Fryer}, C.~L. \& {Heger}, A. 2000, \apj, 541, 1033

\bibitem[{Fryer {et~al.}(2001)Fryer, Woosley, \& Heger}]{fryer01}
Fryer, C.~L., Woosley, S.~E., \& Heger, A. 2001, \apj, 550, 372

\bibitem[Fujimoto et al.(2006)]{fuji06}
Fujimoto, S.-i., Kotake, K., Yamada, S., Hashimoto, M.-a., \& Sato, K.\ 2006, \apj, 644, 1040

\bibitem[{{Glover}(2005)}]{glov05}
{Glover}, S. 2005, \ssr, 117, 445

\bibitem[{{Heger} {et~al.}(2003){Heger}, {Fryer}, {Woosley}, {Langer}, \&
  {Hartmann}}]{hege03}
{Heger}, A., {Fryer}, C.~L., {Woosley}, S.~E., {Langer}, N., \& {Hartmann},
  D.~H. 2003, \apj, 591, 288

\bibitem[{Heger \& Woosley(2002)}]{heger02}
Heger, A. \& Woosley, S.~E. 2002, \apj, 567, 532

\bibitem[{{Iocco} {et~al.}(2005){Iocco}, {Mangano}, {Miele}, {Raffelt}, \&
  {Serpico}}]{iocco05}
{Iocco}, F., {Mangano}, G., {Miele}, G., {Raffelt}, G.~G., \& {Serpico}, P.~D.
  2005, Astroparticle Physics, 23, 303

\bibitem[{{Itoh} {et~al.}(1989){Itoh}, {Adachi}, {Nakagawa}, {Kohyama}, \&
  {Munakata}}]{itoh89}
{Itoh}, N., {Adachi}, T., {Nakagawa}, M., {Kohyama}, Y., \& {Munakata}, H.
  1989, \apj, 339, 354

\bibitem[{Iwamoto {et~al.}(2005)Iwamoto, Umeda, Tominaga, Nomoto, \&
  Maeda}]{iwam05}
Iwamoto, N., Umeda, H., Tominaga, N., Nomoto, K., \& Maeda, K. 2005, Science,
  309, 451

\bibitem[{Kotake {et~al.}(2006)Kotake, Sato, \& Takahashi}]{kotarev}
Kotake, K., Sato, K., \& Takahashi, K. 2006, Rep. Prog. Phys., 69, 971

\bibitem[{Kotake {et~al.}(2004{\natexlab{a}})Kotake, Sawai, Yamada, \&
  Sato}]{kota04a}
Kotake, K., Sawai, H., Yamada, S., \& Sato, K. 2004{\natexlab{a}}, \apj, 608,
  391

\bibitem[{{Kotake} {et~al.}(2003){Kotake}, {Yamada}, \& {Sato}}]{kota03}
{Kotake}, K., {Yamada}, S., \& {Sato}, K. 2003, \prd, 68, 044023

\bibitem[{Kotake {et~al.}(2004{\natexlab{b}})Kotake, Yamada, Sato, Sumiyoshi,
  Ono, \& Suzuki}]{kota04b}
Kotake, K., Yamada, S., Sato, K., Sumiyoshi, K., Ono, H., \& Suzuki, H.
  2004{\natexlab{b}}, \prd, 69, 124004

\bibitem[{{LeBlanc} \& {Wilson}(1970)}]{lebl70}
{LeBlanc}, J.~M. \& {Wilson}, J.~R. 1970, \apj, 161, 541

\bibitem[{{Maki} \& {Susa}(2004)}]{maki04}
{Maki}, H. \& {Susa}, H. 2004, \apj, 609, 467

\bibitem[{Nakamura \& Umemura(2001)}]{naka01}
Nakamura, F. \& Umemura, M. 2001, \apj, 548, 19

\bibitem[{{Nakazato} {et~al.}(2006){Nakazato}, {Sumiyoshi}, \&
  {Yamada}}]{naka06}
{Nakazato}, K., {Sumiyoshi}, K., \& {Yamada}, S. 2006, \apj, 645, 519

\bibitem[{{Obergaulinger} {et~al.}(2006){Obergaulinger}, {Aloy}, \&
  {M{\"u}ller}}]{ober06}
{Obergaulinger}, M., {Aloy}, M.~A., \& {M{\"u}ller}, E. 2006, \aap, 450, 1107

\bibitem[{{Ohkubo} {et~al.}(2006){Ohkubo}, {Umeda}, {Maeda}, {Nomoto},
  {Suzuki}, {Tsuruta}, \& {Rees}}]{ohku06}
{Ohkubo}, T., {Umeda}, H., {Maeda}, K., {Nomoto}, K., {Suzuki}, T., {Tsuruta},
  S., \& {Rees}, M.~J. 2006, \apj, 645, 1352

\bibitem[{{Proga} {et~al.}(2003){Proga}, {MacFadyen}, {Armitage}, \&
  {Begelman}}]{prog03b}
{Proga}, D., {MacFadyen}, A.~I., {Armitage}, P.~J., \& {Begelman}, M.~C. 2003,
  \apjl, 599, L5

\bibitem[{Sandick {et~al.}(2006)Sandick, Olive, Daigne, \& Vangioni}]{sand06}
Sandick, P., Olive, K.~A., Daigne, F., \& Vangioni, E. 2006, \prd, 73,
  104024

\bibitem[{Sawai {et~al.}(2005)Sawai, Kotake, \& Yamada}]{sawa05}
Sawai, H., Kotake, K., \& Yamada, S. 2005, \apj, 631, 446

\bibitem[{Scannapieco {et~al.}(2005)Scannapieco, Madau, Woosley, Heger, \&
  Ferrara}]{scan05}
Scannapieco, E., Madau, P., Woosley, S., Heger, A., \& Ferrara, A. 2005, \apj,
  633, 1031

\bibitem[{{Schneider} {et~al.}(2002){Schneider}, {Guetta}, \&
  {Ferrara}}]{schn02}
{Schneider}, R., {Guetta}, D., \& {Ferrara}, A. 2002, \mnras, 334, 173

\bibitem[{Seto {et~al.}(2001)Seto, Kawamura, \& Nakamura}]{seto01}
Seto, N., Kawamura, S., \& Nakamura, T. 2001, \prl, 87, 221103

\bibitem[{Shen {et~al.}(1998)Shen, Toki, Oyamatsu, \& Sumiyoshi}]{shen98}
Shen, H., Toki, H., Oyamatsu, K., \& Sumiyoshi, K. 1998, \nphysa, 637, 435

\bibitem[{{Stone} \& {Norman}(1992)}]{ston92}
{Stone}, J.~M. \& {Norman}, M.~L. 1992, \apjs, 80, 753

\bibitem[{Takiwaki {et~al.}(2004)Takiwaki, Kotake, Nagataki, \& Sato}]{taki04}
Takiwaki, T., Kotake, K., Nagataki, S., \& Sato, K. 2004, \apj, 616, 1086

\bibitem[{Takiwaki {et~al.}(2007)Takiwaki, Kotake, Yamada, \& Sato}]{taki06}
Takiwaki, T., Kotake, K., Yamada, S., \& Sato, K. 2007, in preparation

\bibitem[{{Umeda} \& {Nomoto}(2002)}]{umed02}
{Umeda}, H. \& {Nomoto}, K. 2002, \apj, 565, 385

\bibitem[{{Umeda} \& {Nomoto}(2003)}]{umed03}
---. 2003, \nat, 422, 871

\bibitem[{Ungarelli {et~al.}(2005)Ungarelli, Corasaniti, Mercer, \&
  Vecchio}]{unga05}
Ungarelli, C., Corasaniti, P., Mercer, R., \& Vecchio, A. 2005, Class. Quant.
  Grav., 22, S955

\bibitem[{Weinmann \& Lilly(2005)}]{wein05}
Weinmann, S.~M. \& Lilly, S.~J. 2005, \apj, 624, 526

\bibitem[{Yamada \& Sawai(2004)}]{yama04}
Yamada, S. \& Sawai, H. 2004, \apj, 608, 907

\end{thebibliography}
\end{document}